# INDIVIDUAL AREA AND SPATIAL DISTRIBUTION OF SAPLINGS

PALAGHIANU C.
*Ştefan cel Mare University of Suceava, Forestry Faculty, Suceava, Romania*
*Correspondig author e-mail address:* cpalaghianu@usv.ro

ABSTRACT: Forest regeneration is a spatial multifaceted process with numerous unknown variables. The individual area or the area potentially available (APA) to an individual plant embodies an idea widely used in population ecology but it has fewer applications in forest research. It was used a Voronoi/ Thiessen tessellation in order to determine the individual area of each sapling. The study was conducted in a naturally regenerated area; using data collected from over seven thousands saplings positioned in a network of permanent rectangular sampling plots. The Voronoi/ Thiessen polygons were used to characterize spatial pattern of sapling distribution as well as the competition relations between the individuals. It is obvious that, at least from the mathematically point of view, the Voronoi tessellation represents one of the best solutions to determine neighbouring competitors of a tree. There were studied the correlations between APA values and the main biometrical attributes, height growth and competition indices. Furthermore, it is shown that APA coefficient of variation is a straight-forward indicator with positive results as an indicator of spatial pattern. The statistical significance of this indicator was evaluated by comparing the results with the values of a 95% confidence envelope generated by Monte-Carlo simulations. Two practical software tools were produced using Visual Basic (VORONOI and ARIA VORONOI) in order to simplify the analyses.

## 1 INTRODUCTION

Researches used many times mathematical and especially geometrical techniques in their effort to explain individual competition.

The area potentially available (APA) concept represents an uncommon, but rather promising approach, introduced in plant ecology by Brown [2]. The same concept was independently developed by Mead [7], but early investigations in the field of plants growing space were conducted also by Konig, mentioned in his book „Die Forst-Mathematik" [6].

From the biological point of view, APA generally defines the area used by an individual to access vital resources, the available area for a plant to satisfy its needs in water, nutrients and light. So APA is very appealing to researchers interested in growth modelling, in their effort to solve an everlasting problem: *"Do trees grow faster because they are larger? Or they are larger because they have been growing faster?"* [17][4].

Considering the difficulty of the analysis there are few researches using this approach [8][9]. Smith [15] considers that this approach is ignored or even avoided due to misapprehend of APA geometrical foundation and computing difficulties. The late period is well-known for its computer development and also numerous and various algorithms were produced. So, the APA re-enters in researcher's attention as a promising investigation tool.

The APA was used to solve not only competition issues but also mortality and dynamics of seedlings [12] or spatial pattern [8]. Regarding spatial pattern, Garcia [4] considers the interaction between neighbouring growing areas as a result of autocorrelation. Two neighbours who are closer than average, will both have APA undersized values and vice versa. Winsauer and Mattson [18] have mentioned some advantages to make use of APA in forest researches – potentially available areas are not intersecting each other, there are sensitive to population dynamics and they are correlated with growth rates. This final remark represents the key aspect of APA utilisation as a competition evaluation tool because if an individual has a large APA, the competition pressure will affect it less. There is, of course, a drawback – the APA is based exclusively on the position of the individual and not on its biometrical attributes. That's why it is called "potentially".

The objective of this study is to elucidate what kind of information APA can offer regarding sapling populations. Can APA characterize the relationships between saplings? A subsidiary objective is producing software tools for Voronoi analysis.

## 2 METHODOLOGY

2.1 What is APA?

The area potentially available of a tree has experienced different forms of interpretation and use, analogous to Brown concept. For example, Staebler [16] Bella [1] and Moore [9] used in their researches a similar concept named "influence zone". Polygon areas were used as descriptive tool of spatial plant arrangement or as predictive tool of plant performance.

The most correct interpretation remains although the one based on the mathematical concept of space partitioning using Voronoi or Thiessen tessellation. So it is generally admitted that APA of an individual is equivalent to area of the Voronoi/ Thiessen polygon which is associated to that individual.

In the bi-dimensional space, a Voronoi polygon of an element includes all the points closer to that specific element than to any other element (Figure 1).

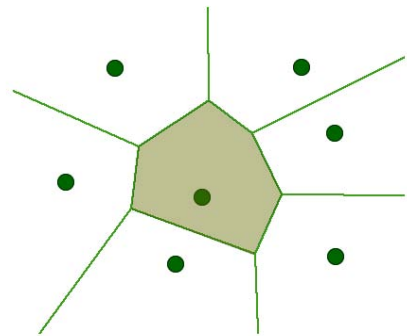

**Figure 1:** Voronoi / Thiessen polygon

The edges of a polygon contain the points located at equal distance from two elements. The vertices of such a polygon are equally located from minimum three generating elements. Considering these properties, two elements are considered to be neighbours if their associated Voronoi polygons share an edge.

It is quite difficult to obtain a Voronoi partitioning for a large set of points, that's why this process is frequently done using specific algorithms and a computer. Algorithms were poorly optimized and the computers were very slow few decades ago, so the process was not pleasant and quick. The last years arrived with great improvements regarding the algorithms and the computer instruments, giving a new chance to Voronoi based applications.

2.2 Developing the software tools

In order to study the area potentially available to saplings I have developed specific software tools, using Microsoft Visual Basic. For the first tool, called VORONOI (Figure 2), I have used an algorithm presented by Ohyama [11] with $O(n^2)$ complexity. VORONOI is stand-alone software which is drawing the Voronoi diagrams using as input data the saplings coordinate placed in a spreadsheet. The user can obtain information regarding sapling neighbours by diagrams analyses. The Voronoi tessellation represents a natural method to select neighbouring trees, a difficult issue in assessing competition indices. The diagrams also offer information about spatial pattern of saplings – it's easier to determine if a pattern is aggregate or uniform.

The second software tool, named ARIA VORONOI, computes the area of each Voronoi polygon. These areas, equivalent to APA values, might be used as competition or aggregation index. Small values of APA might indicate competition pressure and great values of APA coefficient of variation might indicate aggregation of saplings for the analysed plot.

This software was also developed in Microsoft Visual Basic. The input data represents the saplings Cartesian coordinates, extracted from a spreadsheet. The programme computes area of Voronoi polygons and several statistic indicators - the average, standard deviation and coefficient of variation of APA values. It is generated a grid and each cell of the grid is analysed to asses which the generator point (sapling) is. The user can choose a grid size step in order to increase accuracy of determining APA values.

It was taken into account the edge effect, so the saplings with incomplete APA were eliminated from the analyses.

User can specify a value for the buffer zone – in this way the APA is calculated only for the saplings located in the core area, even if the APA extends outside the core area. If the buffer zone is too small, in some exceptional cases, there might be saplings located in the core area with incomplete APA (Figure 3). The algorithm computes also the convex hull and all the points (saplings) located on the convex hull are eliminated. The recommended size of the buffer zone is the average distance between neighbours corrected with the coefficient of variation (20 cm in this study).

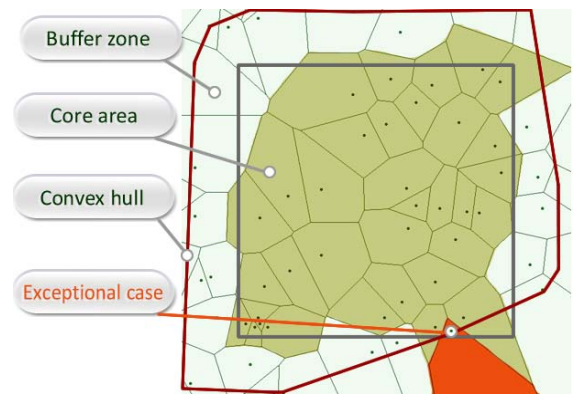

**Figure 3:** Correcting the edge effect

2.3 Material and analyses

The study area is located in Flămânzi Forest District, parcel 50A, near Cotu, a small settlement situated in Botoşani County, Romania. The topography is almost flat, with a slope average of 2-3% and the altitude is around 140 meters. The area of the stand studied is 21.5 hectares and the species composition consists of 30% sessile oak, 20% oak, 30% common hornbeam, 10% small-leaved linden and 10% common ash. The area is regenerated naturally and the regeneration gaps were created in 2001-2002 and were enlarged in 2007. Within this stand a 2.5 hectare homogenous area covered in saplings was selected for further investigation.

I installed a network of ten permanent rectangular sampling plots (7 x 7 m) where I measured the characteristics of all saplings and seedlings (Figure 4).

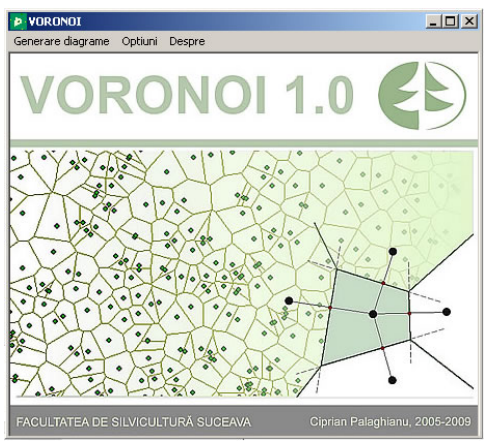

**Figure 2:** Voronoi software interface

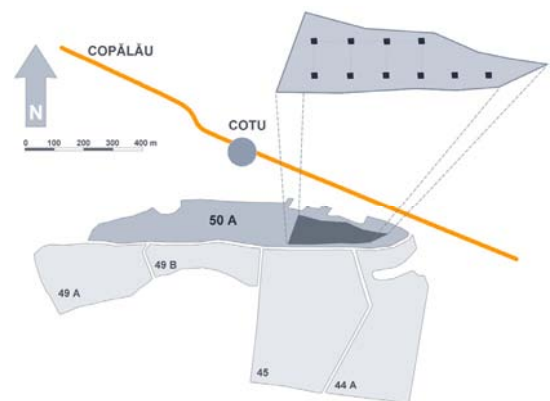

**Figure 4:** Location of the study

I used a GPS receiver in order to record the coordinates of the centre of each plot and I labelled every sapling and seedling inside the plot. The features of 7253 individuals were determined. The attributes assessed are: species, location of the individuals (x, y Cartesian coordinates), diameter, total height, crown insertion height, two crown diameters along the directions of axes and the latest annual height growth.

The analyses were based on areas of generated Voronoi polygons. There were studied the correlations between APA values and the main structural attributes (dimensional attributes, density), height growth and competition indices – Hegyi [5] and Schutz [14].

APA it was also used as a potentially spatial pattern indicator. There are several studies [4][8] that point out there is a relation between APA values or APA coefficient of variation and the spatial pattern of a population of mature trees. This fact might be relevant to sapling populations, too. In this case it was analysed the APA coefficient of variation for each plot in relation with Morisita [10] and Clark-Evans [3] spatial pattern indicators. APA coefficient of variation might be an indicator of spatial distribution. In order to find a relationship between aggregation and APA coefficient of variation values I have used a statistical test to establish if there is a significant deviation from Poisson spatial distribution (from the complete spatial randomness - CSR hypothesis). I have generated 19 Monte-Carlo simulations for each plot, using SpPack software [13] to simulate a CSR distribution for the same area and the same number of saplings. The extreme values of APA coefficient of variation produced the 95% confidence envelope of CSR hypothesis. Higher values of APA coefficient of variations would indicate significant deviations from CSR towards aggregation.

## 3 RESULTS

At first I have studied the relation between APA and the main biometrical attributes. There were identified very significant correlations with low intensity of APA with sapling diameter (r = 0.23***) and crown diameter (r = 0.21***). This is an expected result because several researchers [9], [15], [18] mentioned correlations of mature trees APA with the diameter or basal area.

Obviously there is a strong negative correlation between APA and sapling density (r = - 0.99 ***) and even between APA coefficient of variation and density (r = - 0.72 *). The uniformity tendency is more evident at a higher density.

Several studies [9], [18] indicate that APA might be correlated with growth and competition. Competition is one of the processes that shape the saplings spatial distribution. Consequently there were analysed the correlations between APA and competition indices – Schutz index and Hegyi index computed in respect to diameter, height, crown volume and crown external surface. The strongest correlation is between APA and the Hegyi index calculated in respect to diameter (r = - 0.32 ***) and height (r = - 0.27 ***).

In order to evaluate the performance of growth it was analysed the correlation between APA and sapling height growth. Surprisingly, there is no correlation between these parameters (r = 0.08 *). Other studies indicated significant correlations between mature trees growth (diameter growth) and APA but sapling populations seem to be more dynamic than mature trees. This initial developing stage is very unstable regarding spatial distribution of the individuals so APA is a factor with a smaller impact on height growth.

Some authors [8] point out there is a relation between APA values or APA coefficient of variation and the spatial pattern of a population of mature trees. This seems to be relevant to sapling populations, because of the APA correlations with spatial pattern indicators – Morisita (r = 0.70 *) and Clark-Evans (r = -0.84 **). The APA coefficient of variation is also correlated with spatial pattern indicators Morisita (r = 0.88 **) and Clark-Evans (r = -0.58). Aggregated patterns lead to higher values of APA coefficient of variation (Figure 5 a, b).

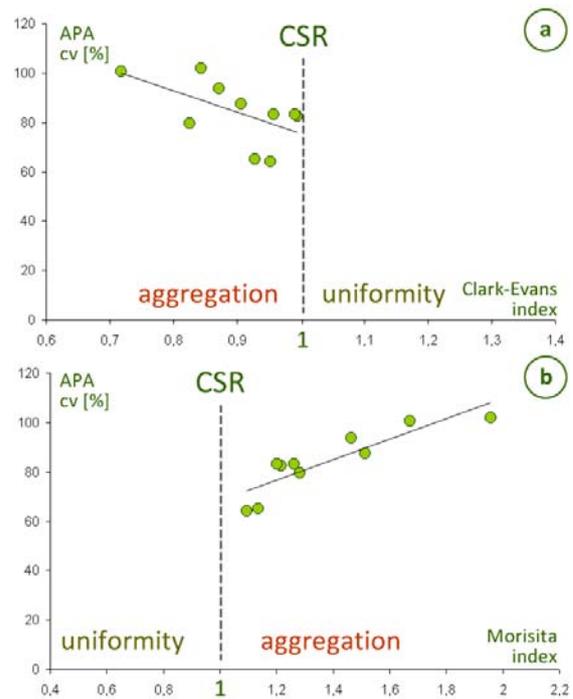

**Figure 5:** APA coefficient of variation related to Clark-Evans (a) and Morisita (b) index

This detail shows that APA coefficient of variation might be used as an indicator of spatial distribution. The Monte-Carlo simulations indicate significant deviations from CSR towards aggregation - the values of APA coefficient of variation overcome the 95% confidence envelope for all the plots (Figure 6).

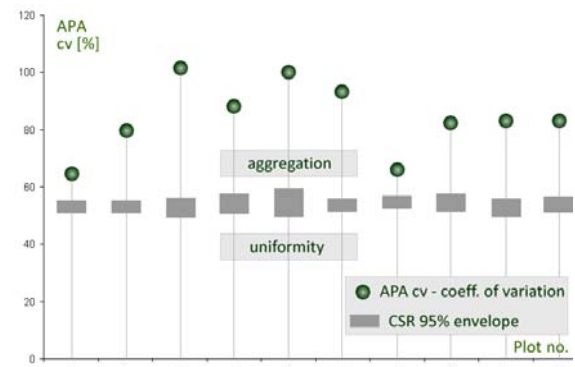

**Figure 6:** APA coefficient of variation values showing aggregation of saplings in all the ten plots

## 4 CONCLUSIONS

The results indicated that saplings APA have poor relationships with biometrical attributes, at least in saplings populations. The reason might be the fact that APA takes into account only sapling position and no other biometrical feature. There is a solution - the generation of Voronoi weighted diagram in respect to some biometric parameter (Figure 7). VORONOI software has the capability to generate such diagrams. Still, there is one problem because it's very difficult to compute the area of the resulted cells.

APA can be described as a low performance indicator of competition in saplings population because there is no relation to height growth and there are low intensity correlations with competition indices. There might be a possibility to use APA in competition analyses in combination with other attributes, but not as a stand-alone indicator. However, one important aspect in assessing competition is that the non-weighted diagrams are the best mathematical solution to establish the neighbours of a sapling or tree. So APA might be used as a criterion for selecting neighbours.

A remarkable result is that APA coefficient of variation represents a straight-forward indicator with positive results as an indicator of spatial pattern. The significance of this indicator might be evaluated by comparing the results with the values of a confidence envelope as it was shown in the paper.

As a final conclusion, APA is a complex and useful tool for characterizing population structure regarding spatial distribution, but seems more suited to mature trees than sapling populations. I hope the development of software tools VORONOI and ARIA VORONOI will simplify and support further studies.

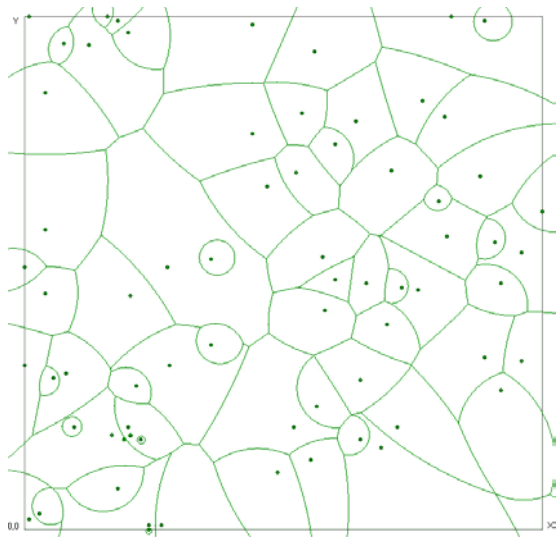

**Figure 7:** Weighted Voronoi diagram in respect to height generated by VORONOI software

## 5 ACKNOWLEDGEMENTS

This work has been mainly supported by the European Commission under the theme "Environment" of the 7th Framework Programme for Research and Technological Development (FP7), Grant agreement no. 226544 (MOTIVE project).

The views expressed herein are exclusively those of the author and do not necessarily reflect those of the European Commission or any other institution or organisation.

## 6 REFERENCES

[1] Bella, I.E., A new competition model for individual trees, Forest Science (1971), 17: 364–372.
[2] Brown, G.S., Point density in stems per acre, New Zealand Forestry Service Research Notes (1965), 38: 1-11.
[3] Clark, P. J., Evans, F. C., Distance to nearest neighbour as a measure of spatial relationships in populations. Journal of Ecology (1954), 35: 445-453.
[4] Garcia, O., Plant individual-based modelling: More than meets the eye, World Conference on Natural Resource Modeling Warsaw (2008), http://forestgrowth.unbc.ca/warsaw.pdf.
[5] Hegyi, F., A simulation for managing jack-pine stands, Growth Models for Tree and Stand Simulation. Royal College of Forestry, Stockholm (1974), Sweden, 74–90.
[6] Konig, G., Die Forst-Mathematik in den Grenzen wirtschaftlicher Anwendung nebst Hülfstafeln für die Forstschätzung und den täglichen Forstdienst, 4th Ed – 1854 [1835], pg. 830.
[7] Mead, R., A relationship between individual plant-spacing and yield. Annals of Botany (1966) 30: 301-309.
[8] Mercier, F., Baujard, O., Voronoi diagrams to model forest dynamics in French Guiana, GeoComputation '97 & SIRC '97 Proceedings, University Otago, New Zealand (1997), 161-171.
[9] Moore, J.A., Budelsky. C.A., Schlesinger, R.C., A new index representing individual tree competitive status, Canadian Journal of Forest Research (1973) 3: 495-500.
[10] Morisita, M., Iδ index, a measure of dispersal of individuals. Researches on Population Ecology (1962) 4: 1-7.
[11] Ohyama, T., Voronoi diagram (2008), http://www.nirarebakun.com/eng.html;
[12] Owens, M.K., Norton, B.E., The impact of 'available area' on Artemisia tridentata seedling dynamics. Vegetatio (1989) 82: 155-162.
[13] Perry, G.L.W., SpPack: spatial point pattern analysis in Excel using Visual Basic for Applications (VBA), Environmental Modelling & Software (2004) 19: 559–569.
[14] Schutz, J.P., Zum Problem der Konkurrenz in Mischbeständen. Schweiz. Z. Forstwes (1989) 140: 1069–1083.
[15] Smith, W.R., Area potentially available to a tree: a research tool, The 19th Southern Forest Tree Improvement Conference (1987), Texas, pg. 29.
[16] Staebler, G.R., Growth and spacing in an even-aged stand of Douglas fir, Master's thesis, University of Michigan (1951), pg. 46.
[17] Wichmann, L., Modelling the efects of competition between individual trees in forest stands, PhD Thesis, The Royal Veterinary and Agricultural University, Copenhagen (2002), pg. 112.
[18] Winsauer, S.A., Mattson, J.A., Calculating Competition In Thinned Northern Hardwoods, Res. Paper NC-306, St. Paul, USDA (1992), pg. 10.